\documentclass{ws-procs10x7}
\usepackage{balance}

\usepackage{amsmath}
\usepackage{amssymb}
\usepackage{amsfonts}
\usepackage{amscd}
\usepackage{bbm} 
\usepackage{mathrsfs}
\usepackage{graphicx}
\usepackage{braket} 
\usepackage{subfigure}
\usepackage{accents}

\def\<{\left\langle}
\def\>{\right\rangle}

\def\D{\mathrm{d}}

\def\be{\begin{equation}}
\def\ee{\end{equation}}
\def\beq{\begin{equation}}
\def\eeq{\end{equation}}
\def\bea{\begin{eqnarray}}
\def\eea{\end{eqnarray}}

\newcommand{\ldm}{\Delta m_{31}^2}

\newcommand{\CenterEps}[2][1]{\ensuremath{\vcenter{\hbox{\includegraphics[scale=#1]{#2.eps}}}}}

\def\<{\left\langle}
\def\>{\right\rangle}

\newcolumntype{d}[1]{D{.}{.}{#1}}

\makeindex

\allowdisplaybreaks[2]

\begin{document}

\title{From Unified Theories to Precision Neutrino Experiments}

\author{Stefan Antusch}

\address{
Departamento de F\'{\i }sica Te\'{o}rica C-XI 
and 
Instituto de F\'{\i }sica Te\'{o}rica C-XVI, \\
Universidad Aut\'{o}noma de Madrid, 
Cantoblanco, E-28049 Madrid, Spain
\\$^*$E-mail: antusch@delta.ft.uam.es}

\twocolumn[\maketitle\abstract{
The expected high sensitivities of future neutrino oscillation experiments will 
allow precision tests of unified theories of flavour. In order to compare 
GUT scale predictions for neutrino masses, leptonic mixing 
angles and CP violating phases with experimental results obtained at low energy, 
their renormalization group running has to be taken into account. 
We discuss the running in seesaw scenarios,  
present analytical approximations as well as numerical tools,   
and compare the size of the running effects with the expected experimental  
sensitivities.  
}
\keywords{Neutrino physics; Unified theories; Renormalization group.}
]

\section{Introduction}
Flavour models within the context of unified theories typically predict 
patterns of fermion masses and mixings at high energies, close to the scale of gauge
coupling unification $M_\mathrm{GUT} \approx 10^{16}$ GeV. 
Our knowledge about neutrino masses and leptonic mixing angles, on the other hand,  
stems from experiments on neutrino oscillations at low energy. 
In order to compare high energy predictions with low 
energy experimental data, the renormalization group (RG) running of neutrino masses, leptonic mixing angles and CP violating phases has to be taken into account.

\section{RG Running in Seesaw Scenarios}
\begin{figure*}
\centerline{\CenterEps[0.8]{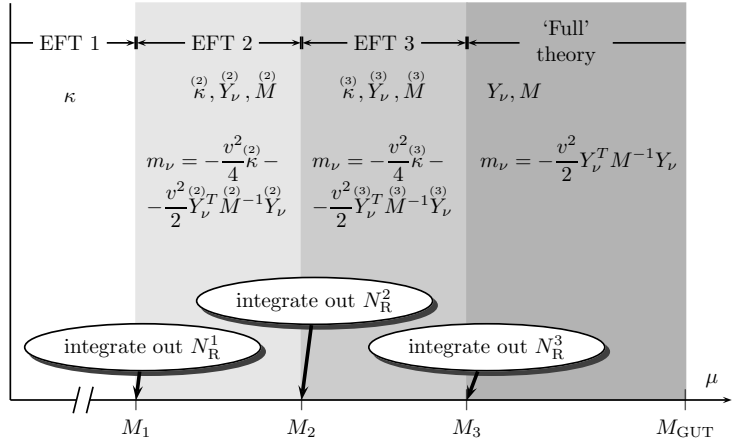}}
\caption{\label{fig2}
Effective theories (EFTs) in minimal seesaw scenarios, where the
heavy right-handed (singlet) neutrinos $N_\mathrm{R}^i$ are partly integrated out. 
At the mass thresholds $M_i$, the EFTs are related by matching conditions. 
 }
\end{figure*}

In seesaw scenarios for explaining the smallness of neutrino masses, 
the strategy is to successively solve the systems
of coupled differential equations of the form
\begin{eqnarray}
\mu \frac{\D}{\D \mu}   \accentset{(n)}{X}_i
  = \accentset{(n)}{\beta}_{{X}_i} (\{
  \accentset{(n)}{X}_j\})
\end{eqnarray}
for all the parameters $\accentset{(n)}{X}_i \in \{\accentset{(n)}{\kappa},\accentset{(n)}{Y_\nu},\accentset{(n)}{M},
\dots\}$
of the theory, including the effective dimension 5 neutrino mass operator $\kappa$, the neutrino Yukawa matrix $Y_\nu$ and the mass matrix $M$ of the right-handed (singlet) neutrinos, as illustrated in Fig.~\ref{fig2}.
The corresponding parameters defined in 
the energy ranges corresponding to the various effective 
theories are marked by $(n)$. 

Below the lowest mass threshold, the RGEs (given for various models in Refs.~\refcite{b1} to \refcite{Grimus:2004yh}) of the dimension 5 neutrino mass operator can be used.
In the MSSM, it is known at two-loop.\cite{2loop}
For non-degenerate seesaw scales, a method for dealing with the effective
theories where the
heavy singlets are partly integrated out, and the corresponding RGEs, 
can be found in Ref.~\refcite{SeesawRunning}.

\section{REAP/MPT: Public Software Packages for RG Running}
For RG running from high energy (e.g.\ from the GUT scale) to low energy in the above-described minimal 
seesaw scenarios, we provide a public Mathematica software 
package REAP (Renormalization Group Evolution
of Angles and Phases), introduced in Ref.~\refcite{Antusch:2005gp}, which numerically solves the RGEs of the quantities relevant
for neutrino masses, for example the dimension 5 neutrino mass operator, the
Yukawa matrices and the gauge couplings.  
The $\beta$-functions for the SM, the
MSSM and two Higgs doublet models with $\mathbb{Z}_2$ symmetry for FCNC
suppression, with and without right-handed neutrinos, are implemented.  In
addition, the same models are available for pure Dirac neutrinos,\cite{Lindner:2005as} and new models can be
added by the user. The software can also be applied to type II seesaw
models\cite{Antusch:2004xd} with an additional contribution to the neutrino mass matrix from heavy SU(2)$_{\mathrm{L}}$-triplet Higgs fields $\Delta_\mathrm{L}$, as long as one only considers the energy region below the additional
seesaw scale $M_\Delta$, where the new physics only
leads to another contribution to the effective neutrino mass operator. 
The package can be downloaded from http://www.ph.tum.de/\~{}rge/REAP/.

Combined with the additionally provided public Mathematica package MPT (MixingParameterTools),  
which allows to extract the
masses, mixing angles and CP phases of quarks and leptons from their 
mass matrices, the running of the neutrino parameters can be 
calculated conveniently.

\section{Analytical Formulae for the Running of the Parameters}
Below the seesaw scales, up to ${\mathcal{O}} (\theta_{13})$ corrections, the evolution of the
mixing angles is given by\cite{Antusch:2003kp}
(see also Refs.~\refcite{Chankowski:1999xc} and \refcite{Casas:1999tg}) 
\begin{eqnarray}
\label{Eq:t12}\dot{\theta}_{12} \!&=&\!
        \frac{- C_e y_\tau^2}{32\pi^2} \,
        \!\sin 2\theta_{12} \, s_{23}^2\,  
      \frac{
      | {m_1}\, e^{i \varphi_1} \!\!+\! {m_2}\, e^{i  \varphi_2}|^2
     }{\Delta m^2_{\odot} },\nonumber\\
     &&
\\ \nonumber
\label{Eq:t13}\dot{\theta}_{13} \!&=&\!
        \frac{C_e y_\tau^2}{32\pi^2} \,
        \sin 2\theta_{12} \, \sin 2\theta_{23} \,
        \frac{m_3}{\Delta m^2_{\rm A} \left( 1+\zeta \right)} \nonumber \\
      &&  \times \: I(m_i, \varphi_i, \delta)
\,,
\\ \nonumber
\label{Eq:t23}     \dot{\theta}_{23} \!&=&\! 
        \frac{-C_e y_\tau^2}{32\pi^2} \,
        \frac{\sin 2\theta_{23}}{\Delta m^2_{\rm A}}
       \left[
         c_{12}^2 \, |m_2\, e^{i \varphi_2} + m_3|^2   \right.      \nonumber \\
      && + \!\left. s_{12}^2 \, \frac{|m_1\, e^{i \varphi_1}\! + m_3|^2}{1+\zeta}
        \right],
\label{eq:Theta23Dot}
\end{eqnarray}
where the dot indicates differentiation $d/dt = \mu\,d/d\mu$ (with $\mu$ being
the renormalization scale), and where $s_{ij}:=\sin\theta_{ij}$, 
$c_{ij} = \cos\theta_{ij}$, $\zeta = \Delta m^2_{21}/\Delta m^2_{31}$,
$C_e = -3/2$ in the SM and $C_e = 1$ in the MSSM, and   
\begin{eqnarray}
\lefteqn{I(m_i, \varphi_i,\delta)
=\left[ m_1 \cos(\varphi_1-\delta) \right.}\nonumber \\
&& \left.  
- ( 1\!+\!\zeta ) m_2 
\cos(\varphi_2-\delta) - \zeta m_3  \cos\delta \right].
\end{eqnarray} 
$y_\tau$ denotes the tau Yukawa coupling, and one can safely neglect the contributions coming from the electron and muon. For the matrix $P$ containing the Majorana phases,
we use the convention $P = \mbox{diag}\, (e^{-i \varphi_1/2},e^{-i
\varphi_2/2},1)$ as in Ref.~\refcite{Antusch:2003kp}. From these expressions one can easily understand the typical
size of RG effects as well as some basic properties:
\begin{itemize}
\item Below the seesaw scales, the RGEs are proportional to $y^2_\tau$. In the MSSM, $y_\tau$ can be ${\cal O}(1)$ for large $\tan \beta$ and the running is enhanced by a factor of $(1+\tan^2 \beta)$.

Note that above and between the seesaw scales, additional terms appear, e.g.\ proportional to $y^2_\nu$, with $y_\nu$ being a neutrino Yukawa coupling.\cite{Antusch:2005gp} 
They can induce sizable running even for small $y_\tau$. 
\item Due to the terms of the form $m_i/\Delta m_{ij}^2$, 
the running can be strongly enhanced if the neutrino masses have a quasi-degenerate spectrum, with masses much larger than mass splittings.
\item The CP violating phases $\delta$, $\varphi_1$ and $\varphi_2$ 
can either damp or enhance the running, as can be deduced from Eqs.~(\ref{Eq:t12}) - (\ref{Eq:t23}).
\end{itemize}
In addition to the formulae for the running of the mixing angles,  formulae for the running of the CP phases have been derived.\cite{Antusch:2003kp} For example, the running of the Dirac CP violating phase 
$\delta$, observable neutrino oscillation experiments, is given
by        
\begin{equation} \label{eq:DeltaPrimeWithNonZeroDelta}
 \Dot{\delta}
 \,=\,
 \frac{C y_\tau^2}{32\pi^2}
 \frac{\delta^{(-1)}}{\theta_{13}}
 +\frac{C y_\tau^2}{8\pi^2}\delta^{(0)}+\mathscr{O}(\theta_{13})\;.
\end{equation}
The coefficients $\delta^{(-1)}$ and $\delta^{(0)}$ are omitted here for brevity and are given explicitly in Ref.~\refcite{Antusch:2003kp}. From 
Eq.~(\ref{eq:DeltaPrimeWithNonZeroDelta}), it can be seen that the 
Dirac CP phase generically becomes more unstable under RG corrections for smaller  $\theta_{13}$.  

Analytical formulae for the running of the neutrino parameters above the seesaw scales can be found in Ref.~\refcite{Antusch:2005gp} (see also Ref.~\refcite{Mei:2005qp}).

\section{Precision Tests of Unified Flavour Model Predictions}
Future reactor and long-baseline experiments have the potential to measure 
the neutrino parameters with high precision.  
For testing predictions of unified flavour models using such future precision measurements, RG corrections have to be included, as we will now discuss.\cite{Mohapatra:2006gs} 

 For roughly estimating the size of the RG effects, and for identifying cases where RG corrections are enhanced or suppressed, 
 analytical formulae, as those presented above, are very useful. 
In the leading logarithmic approximation, a rough estimate of the corrections to the leptonic mixing angles from RG running between electroweak scale $M_\mathrm{EW}$ and high energy scale $M_\mathrm{GUT}$ can be obtained as  
 $\Delta \theta_{ij} \approx \dot\theta_{ij} \ln (M_\mathrm{GUT}/M_\mathrm{EW})$. 

 For a more accurate numerical calculation of
 the RG running in seesaw models, the public  
 software package REAP, introduced in Ref.~\refcite{Antusch:2005gp}, can be used. 
 The right-handed (singlet) neutrinos are integrated out successively at their mass thresholds and a  
 template is provided for the user (in form of a Mathematica notebook) where unified models of flavour can be inserted conveniently.

\subsection{Running of $\theta_{13}$ and $\delta$}
The mixing angle $\theta_{13}$ is a main focus of the currently  planned neutrino oscillation experiments. From the present experimental data it is only bounded from above by $\approx 13^\circ$, however sensitivity to $\theta_{13}$ less than $1^\circ$ is envisioned by neutrino factory\cite{Blondel:2006su} and/or beta
beam\cite{Burguet-Castell:2005pa} facilities. 
If $\theta_{13}$ is not too small, the Dirac CP phase $\delta$ can be measured 
and a precise determination of $\theta_{13}$ can be achieved.   
Furthermore, $\theta_{13}$ is a good discriminator between models of neutrino masses.\cite{Albright:2006cw}

\begin{figure}
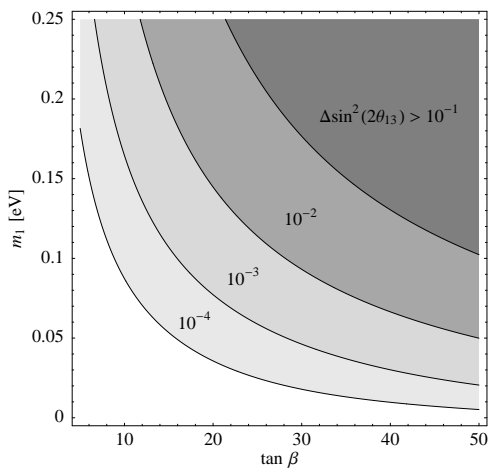

\CenterEps[0.57]{RGCorrectionTheta13}
\caption{\label{fig:RGCorrectionTheta13}
Conservative estimate of RG corrections to $\sin^2(2\theta_{13})$ in the MSSM with $\theta_{13} = 0$ at high energy $10^{12}$ GeV.$^{11}$  
For comparison: Planned reactor and long-baseline experiments will be sensitive to 
$\sin^2(2\theta_{13}) = {\cal O}(10^{-2})$, future upgraded superbeam experiments up to ${\cal O}(10^{-3})$, and neutrino factories and/or beta
beam facilities can reach sensitivities of $\sin^2(2\theta_{13}) = {\cal O}(10^{-4})$, or even better. 
}
\end{figure}

Due to RG running, the low scale value of $\theta_{13}$ differs from its high scale prediction, unless $m_3 = 0$ (and $\theta_{13}=0$) for inverse neutrino mass hierarchy, 
or unless the masses and CP phases are aligned in a special way (c.f.\ Eq.~(\ref{Eq:t13})).  
One consequence is that even if $\theta_{13}=0$ is predicted by some model at high energy, RG running leads to $\theta_{13}\not=0$ at low energy.\cite{Antusch:2003kp,Mei:2004rn}
In many cases, the radiatively generated $\theta_{13}$ is within reach of future experiments. 
This is illustrated in Fig.~\ref{fig:RGCorrectionTheta13} (from Ref.~\refcite{Antusch:2003kp}), where a conservative estimate for the size of the 
radiatively generated $\theta_{13}$ is given for the MSSM with a normal neutrino mass ordering, as a function of $\tan \beta$ and the mass of the lightest neutrino $m_1$. 
 
\vspace{-0.2cm}
As can be seen from Eq.~(\ref{eq:DeltaPrimeWithNonZeroDelta}), the running of the Dirac CP phase $\delta$ is larger for smaller $\theta_{13}$. Even if $\delta = 0$ holds at high energy, RG running can induce a non-zero value at low energy (in the presence of non-zero Majorana CP phases).\cite{Casas:1999tg,Antusch:2003kp} 
In the extreme case, for $\theta_{13}=0$, $\delta$ is undefined. It is then entirely determined by RG running (which also induce a non-zero $\theta_{13}$) and given  by\cite{Antusch:2003kp}
\begin{equation}\label{eq:DeltaZeroT13}
        \cot\delta \!=\!
        \frac{ m_1 \cos\varphi_1 - \left(1+\zeta\right) m_2 \cos\varphi_2-
         \zeta m_3 }
        { m_1 \sin\varphi_1 - \left(1+\zeta\right) m_2 \, \sin\varphi_2 }
        \;.\nonumber 
\end{equation} 
Note that for very small $\theta_{13}$, the equations for $\theta_{13}$ and $\delta$ have to be solved simultaneously.

\subsection{Predictions for $\theta_{12}$}
Many classes of promising unified flavour models lead to predictions for the neutrino mixing angles, such as, for instance, tri-bimaximal\cite{tribi} or bimaximal\cite{bimax} mixing. 
In the presence of additional mixing in the charged lepton mass matrix, however, these mixing patterns are modified. Nevertheless, if the charged lepton mixing matrix is CKM-like, i.e.\ dominated by a 1-2 mixing like the CKM mixing matrix of the quarks, interesting predictions arise for the observable mixing angle $\theta_{12}$ of the MNS matrix. 

Remarkably, it can be shown that under the above conditions, a combination of the measurable MNS parameters $\theta_{12}$, $\theta_{13}$ and $\delta$ sums up to the theoretically predicted value for the 1-2 mixing of the neutrino mass matrix\cite{sumrulesSO(3),sumrules}, i.e.\ to $\arcsin(\tfrac{1}{\sqrt{3}})$ for
tri-bimaximal and $\tfrac{\pi}{4}$ for bimaximal mixing, for example.
The neutrino sum rules with theory predictions of tri-bimaximal and bimaximal neutrino mixing, respectively, are\cite{sumrulesSO(3),sumrules}
\begin{eqnarray}
\label{sumrules1}\theta_{12} - \theta_{13}\cos (\delta) &\approx&  \arcsin \tfrac{1}{\sqrt{3}}\;,\\
\label{sumrules2}\theta_{12} - \theta_{13}\cos (\delta) &\approx&  \tfrac{\pi}{4} \;. 
\end{eqnarray}
By measuring the left side of the neutrino sumrules in future neutrino experiments, and by taking into account RG running of the quantities, 
one can test whole classes of unified flavour models.   

In principle, the so-called quark-lepton complementarity\cite{Raidal:2004iw} (QLC) relation $\theta_{12} + \theta_C = \pi/4$ 
between the leptonic mixing angle $\theta_{12}$ and the Cabibbo angle $\theta_C$ can be obtained within this scenario. However, it requires $\theta_{13} = \theta_\mathrm{C}$, CP conserving $\delta = \pi$ and bimaximal mixing of the neutrino mass matrix, as can be seen from Eq.~(\ref{sumrules2}). 
If QLC is realized in a unified flavour model at the GUT scale, the predicted mixing angles at low energy are modified due to RG running. For instance, within the approach of Ref.~\refcite{Antusch:2005ca} to QLC in unified theories, the low energy value of $\theta_{12}$ is reduced due to running by $\approx 0.8^\circ$ (for $\tan \beta = 40$, calculated with REAP\cite{Antusch:2005gp}). 
RG corrections in various scenarios of QLC have been studied in detail in Ref.~\refcite{Schmidt:2006rb}. 
For further recent studies on RG running of neutrino parameters, see, for example,  Ref.~\refcite{recent}. 
A precise measurement of $\theta_{12}$ could be achieved by a SPMIN-type reactor experiment.\cite{Bandyopadhyay:2004cp}

\subsection{Corrections to Maximal Mixing $\theta_{23}$} 
The present best-fit value of $\theta_{23}$ is close to maximal. Typically, 
unified flavour models predict deviations of $\theta_{23}$ from maximality,  
which are within reach of future long baseline experiments.\cite{Antusch:2004yx}
$\theta_{23}$ close to maximal, on the other hand, would point towards a symmetry which
predicts maximal mixing. 
However, even if $\theta_{23}=\tfrac{\pi}{4}$ is predicted by a flavour model at high energy, RG corrections
from the running between high and low energy generate a deviation
of $\theta_{23}$ from maximality. 
In many cases, even for hierarchical neutrino masses, 
this deviation exceeds the sensitivities of future 
experiments.\cite{Antusch:2003kp,Antusch:2004yx} 
From maximal mixing at $M_\mathrm{GUT}$, the running in the MSSM leads to low energy values $> \tfrac{\pi}{4}$ ($< \tfrac{\pi}{4}$) for normal (inverse) neutrino mass ordering, as can be seen from Eq.~(\ref{Eq:t23}). Note that $\dot \theta_{23}$ is always nonzero in the SM and MSSM, i.e.\ maximal mixing is always unstable.

For instance, assuming a (normal) hierarchical spectrum of light neutrinos and neutrino Yukawa couplings of ${\cal O}(1)$, RG running in the MSSM leads to a deviation from maximal mixing of about $1^\circ$ for $\tan \beta = 10$ up to $5^\circ$ for $\tan \beta = 55$ (using REAP\cite{Antusch:2005gp}). 
The latter corresponds to $\sin^2 (\theta_{23}) \approx 0.59$ at low energy if maximal mixing (i.e.\ $\sin^2 (\theta_{23})= 0.5$) was predicted at the GUT scale. 

\begin{figure}
\CenterEps[1.26]{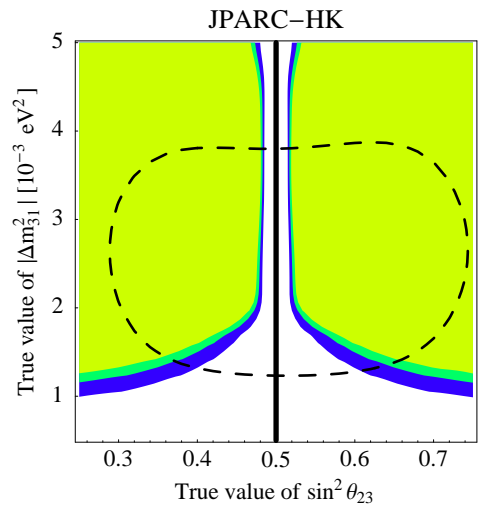}
\caption{\label{fig:T2HK}
The regions of the true values of $\sin^2
\theta_{23}$ and $| \ldm |$ where maximal mixing can be rejected by
 at $1 \sigma$, $2 \sigma$ and $3 \sigma$
JPARC-HK (nowadays referred to as T2HK), from dark to light shading (for details, see 
Ref.~29).  
The currently allowed region at the $3 \sigma$ CL is shown as a dashed curve.
For comparison: RG running in the MSSM with $\sin^2 (\theta_{23}) = 0.5$ at the GUT scale leads to $\sin^2 (\theta_{23}) \approx 0.52$ for $\tan \beta = 10$, 
up to $\sin^2 (\theta_{23}) \approx 0.59$ for $\tan \beta = 55$, in an example with ${\cal O}(1)$ neutrino Yukawa couplings and normal hierarchy for the light neutrino masses ($m_1 \approx 0$).$^8$
}
\end{figure}

Measuring deviations $|\sin^2 (\theta_{23}) - 0.5|$ from maximal mixing is  challenging for the currently planned experiments.\cite{Antusch:2004yx} 
A very sensitive experiment for detecting such deviations would be an upgraded superbeam pointing at a large (1 Mt) water Cherenkov detector, e.g.\ an experiment like T2HK. 
Its potential for rejecting maximal mixing is illustrated in Fig.~\ref{fig:T2HK} (from Ref.~\refcite{Antusch:2004yx}). With such a detector, atmospheric neutrino experiments with high statistics could also provide interesting additional  information.\cite{Gonzalez-Garcia:2004cu}     

\vspace{-0.1cm}
\vspace{-0.6cm}

\section{Conclusions}
Neutrinos, with the smallness of their masses explained by the seesaw mechanism, 
provide an intriguing window to physics at very high energies, 
close to the GUT scale. 
One interesting aspect is that the expected high sensitivities of future neutrino experiments will allow precision tests of unified theories of flavour.
In addition to the high experimental sensitivities, such precision tests also require
high accuracy of model predictions, in particular the inclusion of renormalization group   running of the neutrino parameters.

\section*{Acknowledgments}
I would like to thank 
M.~Drees, P.~Huber, J.~Kersten, M.~Lindner, M.~Ratz, 
M.~A.~Schmidt, T.~Schwetz and W.~Winter 
for the collaboration in the studies presented here. 
I would also like to thank the organizers of 
ICHEP 2006 and I acknowledge supported by the EU 6$^\text{th}$
Framework Program MRTN-CT-2004-503369 ``The Quest for Unification:
Theory Confronts Experiment''.


\begin{thebibliography}{9}




\bibitem{b1}
P.~H.~Chankowski and Z.~Pluciennik,
{\it Phys.\ Lett.}~{\bf B316} (1993) 312.

\bibitem{b2}
K.~S.~Babu, C.~N.~Leung and J.~Pantaleone,
{\it Phys.\ Lett.}~{\bf B319} (1993) 191.

\bibitem{adklr1}
S.~Antusch, M.~Drees, J.~Kersten, M.~Lindner and M.~Ratz,
{\it Phys.\ Lett.}~{\bf B519} (2001) 238, 

\bibitem{adklr2}
S.~Antusch, M.~Drees, J.~Kersten, M.~Lindner and M.~Ratz,
{\it Phys.\ Lett.}~{\bf B525} (2002) 130.

\bibitem{2loop}
S.~Antusch and M.~Ratz,
{\it JHEP} {\bf 0207} (2002) 059;

\bibitem{SeesawRunning}
S.~Antusch, J.~Kersten, M.~Lindner and M.~Ratz,
{\it Phys.\ Lett.}~{\bf B538}~(2002)~87.

\bibitem{Grimus:2004yh}
W.~Grimus and L.~Lavoura, ]
 Eur.\ Phys.\ J.\ C {\bf 39} (2005) 219. 

\bibitem{Antusch:2005gp}
  S.~Antusch, J.~Kersten, M.~Lindner, M.~Ratz and M.~A.~Schmidt,
  JHEP {\bf 0503} (2005) 024.


\bibitem{Lindner:2005as}
  For an analysis of RG running with Dirac neutrino masses, see: 
  M.~Lindner, M.~Ratz and M.~A.~Schmidt,
  JHEP {\bf 0509} (2005) 081.
  
\bibitem{Antusch:2004xd}
  For RG effects in classes type II seesaw models with partially degenerate 
  neutrino masses, see: S.~Antusch and S.~F.~King,
  Nucl.\ Phys.\ B {\bf 705} (2005) 239.  

\bibitem{Antusch:2003kp}
S.~Antusch, J.~Kersten, M.~Lindner, M.~Ratz,
{\it Nucl.\ Phys.}~{\bf B674} (2003) 401.

\bibitem{Chankowski:1999xc}
P.~H. Chankowski, W.~Krolikowski, S.~Pokorski, 
{\it Phys.~Lett.}~\textbf{B473}~(2000)~109.

\bibitem{Casas:1999tg}
J.~A. Casas et al., {\it Nucl.\ 
Phys.}~\textbf{B573} (2000), 652.

\bibitem{Mei:2005qp}
  J.~w.~Mei,
  Phys.\ Rev.\ D {\bf 71} (2005) 073012.

\bibitem{Mohapatra:2006gs}
  For reviews of neutrino physics and models of neutrino masses and mixings, see e.g.: 
  S.~F.~King,
  Rept.\ Prog.\ Phys.\  {\bf 67} (2004) 107;
   G.~Altarelli and F.~Feruglio,
  New J.\ Phys.\  {\bf 6} (2004) 106. 
   R.~N.~Mohapatra {\it et al.} (Theory working group of the APS Neutrino Study),
  [arXiv:hep-ph/0412099] and     
  [arXiv:hep-ph/0510213];
  R.~N.~Mohapatra and A.~Y.~Smirnov,
  [arXiv:hep-ph/0603118];   
  A.~Strumia and F.~Vissani,
  [arXiv:hep-ph/0606054]; 
 J.~W.~F.\ Valle,
  [arXiv:hep-ph/0608101].



 \bibitem{Blondel:2006su}
  A.~Blondel, A.~Cervera-Villanueva, A.~Donini, P.~Huber, M.~Mezzetto and P.~Strolin,
  [arXiv:hep-ph/0606111];
  P.~Huber, M.~Lindner, M.~Rolinec and W.~Winter,
  [arXiv:hep-ph/0606119].

\bibitem{Burguet-Castell:2005pa}
  J.~Burguet-Castell, D.~Casper, E.~Couce, J.~J.~Gomez-Cadenas and P.~Hernandez,
  Nucl.\ Phys.\ B {\bf 725} (2005) 306;
  J.~E.~Campagne, M.~Maltoni, M.~Mezzetto and T.~Schwetz,
  [arXiv:hep-ph/0603172].

\bibitem{Albright:2006cw}
  C.~H.~Albright and M.~C.~Chen,
  [arXiv:hep-ph/0608137]. 

\bibitem{Mei:2004rn}
J.~w.~Mei and Z.~z.~Xing,
Phys.\ Rev.\ D {\bf 70}, 053002 (2004).


\bibitem{tribi}
P.~F.~Harrison, D.~H.~Perkins and W.~G.~Scott,
Phys.\ Lett.\ B {\bf 530} (2002) 167.

\bibitem{bimax}
  V.~D.~Barger, S.~Pakvasa, T.~J.~Weiler and K.~Whisnant,
  Phys.\ Lett.\ B {\bf 437} (1998) 107.



\bibitem{sumrulesSO(3)}
  For a sumrule in a unified flavour model with tri-bimaximal neutrino mixing, see: 
  S.~F.~King,
  JHEP {\bf 0508} (2005) 105;

\bibitem{sumrules}  
  Independent of a specific model, neutrino sumrules are derived in: 
  S.~Antusch and S.~F.~King,
  Phys.\ Lett.\ B {\bf 631} (2005) 42.

%
\bibitem{Raidal:2004iw}
M.~Raidal,
Phys.\ Rev.\ Lett.\  {\bf 93}, 161801 (2004); 
H.~Minakata and A.~Y.~Smirnov,
Phys.\ Rev.\ D {\bf 70}, 073009 (2004).

\bibitem{Antusch:2005ca}
  S.~Antusch, S.~F.~King and R.~N.~Mohapatra,
  Phys.\ Lett.\ B {\bf 618} (2005) 150.

\bibitem{Schmidt:2006rb}
  M.~A.~Schmidt and A.~Y.~Smirnov,
  [arXiv:hep-ph/0607232];

\bibitem{recent}
  T.~Dent, G.~K.~Leontaris, A.~Psallidas and J.~Rizos,
  arXiv:hep-ph/0603228;    
  S.~Luo and Z.~z.~Xing,
   Phys.\ Lett.\ B {\bf 632} (2006) 341; 
  S.~K.~Kang, C.~S.~Kim and J.~Lee,
  Phys.\ Lett.\ B {\bf 619} (2005) 129;
  J.~R.~Ellis, A.~Hektor, M.~Kadastik, K.~Kannike and M.~Raidal,
  Phys.\ Lett.\ B {\bf 631} (2005) 32;
 A.~Broncano, M.~B.~Gavela and E.~Jenkins,
 Nucl.\ Phys.\ B {\bf 705} (2005) 269.   


\bibitem{Bandyopadhyay:2004cp}
  See e.g.: A.~Bandyopadhyay, S.~Choubey, S.~Goswami and S.~T.~Petcov,
  Phys.\ Rev.\ D {\bf 72} (2005) 033013.

\bibitem{Antusch:2004yx}
  S.\ Antusch, P.\ Huber, J.\ Kersten, T.\ Schwetz, W.\ Winter,
Phys.\ Rev.\ D {\bf 70} (2004) 097302.


\bibitem{Gonzalez-Garcia:2004cu}
  M.~C.~Gonzalez-Garcia, M.~Maltoni and A.~Y.~Smirnov,
  Phys.\ Rev.\ D {\bf 70} (2004) 093005; 
    S.~Choubey and P.~Roy,
  Phys.\ Rev.\ D {\bf 73} (2006) 013006;
 S.~T.~Petcov and T.~Schwetz,
  Nucl.\ Phys.\ B {\bf 740} (2006) 1.

\end{thebibliography}
\end{document}